# Using a Stifneck Select Collar™ for hands-free semiautomatic blood flow measurements: a user study

Reinhard Fuchs, Nathalie Sumrah, Maximilian N. Möbius-Winkler, Georg Stachel, Michael Schultz, Ulrich Laufs, Thomas Neumuth, *Member, IEEE*, Michael Unger, and Karsten Lenk

*Abstract*— ***Objective:*** **The percentage of long-term survival in out-of-hospital cardiac arrest cases is remarkably low. One approach would be to increase the effectiveness of cardiopulmonary resuscitation (CPR), which is currently not measurable in a quantifiable way. The most significant challenge in providing a mobile solution for CPR evaluation is a mobile, hazard-free sensor attachment with high usability.** ***Methods:*** **We present a sensor attachment solution usable for semiautomatic ultrasonic (US) Doppler measurements. Components are attached to a Stifneck Select Collar™ (Laerdal). An inflatable cushion (TR-Band™, Terumo) allows adjustable contact pressure. A clinical study was conducted in which the system was evaluated based on comfort, pain, sensor support, the viability of Doppler signals, and the absence of skin irritations.** ***Results:*** **The system was utilized in a prospective study involving 102 healthy probands. On a scale between 1 (Low) and 10 (Intense), ratings were 1.19 (SD 0.46), 6.52 (SD 1.78), and 9.95 (SD 0.32) for pain, comfort, and support, respectively. The average duration of application was 31.19 minutes (SD 16.75 minutes). Audible Doppler signals were achieved in 92.2 % of the probands, and Doppler curve evaluation was usable in 73.5 %. No skin irritations were observed.** ***Conclusion:*** **A hands-free sensor attachment for a US probe was developed that caused no significant complaints by healthy study volunteers. Medical users assessed its attachment as robust.** ***Significance*****: With its adjustable positioning and easy attachment, the Stifneck modification can form a basis for a mobile US Doppler device, capable of evaluating carotid artery flow during CPR.**

This work was supported by the German Federal Ministry for Economic Affairs and Energy under Grant 16KN083328, the Deutsche Herzstiftung e.V., Frankfurt a. M., and the German Federal Ministry of Research, Technology, and Space under Grant 13GW0768B. The publication was supported by the Open Access Publishing Fund of Leipzig University (Michael Unger and Karsten Lenk are co-senior authors.) (Corresponding author: Reinhard Fuchs).

This work involved human subjects or animals in its research. Approval of all ethical and experimental procedures and protocols was granted by ethics committee of the University of Leipzig, Faculty of Medicine (reference number 183/23-ek).

N. Sumrah, M. N. M. Winkler, G. Stachel, U. Laufs, and K. Lenk are with the Klinik und Poliklinik für Kardiologie, Universitätsklinikum Leipzig, Germany.

M. Schultz is with the GAMPT mbH, Merseburg, Germany.

R. Fuchs ,T. Neumuth, and M. Unger are with the Innovation Center Computer Assisted Surgery (ICCAS), Medical Faculty, Leipzig University, 04103 Leipzig, Germany (e-mail: reinhard.fuchs@medizin.uni-leipzig.de).

## I. Introduction

THE survival rates of out-of-hospital cardiac arrest (OHCA) remain low, ranging from 0% to 15% at the time of hospital discharge [1]. Long *no flow time* (NFT) has been identified as one of the crucial factors influencing long-term hospital outcomes after OHCA.

The CPR quality assurance is currently not addressed by modern technologies, since CPR results cannot be evaluated outside of a hospital without extensive medical knowledge and professional tools. Guy et al. estimate that each minute of NFT decreases the favourable neurological outcome after ROSC by 13 %, while Contri et al. state that 10 minutes of NFT allow for the prediction of a 0 % survival chance [2], [3]. Consequently, first responders and medical personnel might therefore profit from a tool capable of automatically detecting and quantifying the blood supply of the common carotid artery (CCA). The device should have the potential to act as a surrogate for cerebral blood flow and could inform about the effectiveness of CPR in real-time, provide records about NFT phases, allow for new insights about long term consequences based on CPR quality, and improve existing guidelines or scoring systems [4], [5], [6], [7], [8], [9], [10], [11], [12], [13], [14], [15], [16], [17].

Non-invasive blood flow measurements during resuscitation are technically feasible with commercially available ultrasound (US) devices. However, this technology has yet to be effectively implemented in pre-hospital situations [6], [7]. While portable ultrasound devices with integrated Power Doppler (e.g. Clarius L7 Linear Handheld Ultrasound Scanner, Siemens ACUSON Freestyle) are available, they cannot be used hands-free [18], [19]. The sensor fixation of the device must be precise to enable the focus of an ultrasonic sample volume (SV) to actually align with the common carotid artery. The sensor also needs sufficient skin contact, typically facilitated by applying a certain degree of pressure. Due to the ongoing CPR, multiple forceful impulses will travel from the chest region throughout the body, which influences sensor attachment and alignment with the target vessel. Any method of sensor placement needs to withstand this kind of forces or the resulting tissue movement. The attachment method must also not interfere with the workflow of resuscitation, featuring high usability and minimal time duration for application.

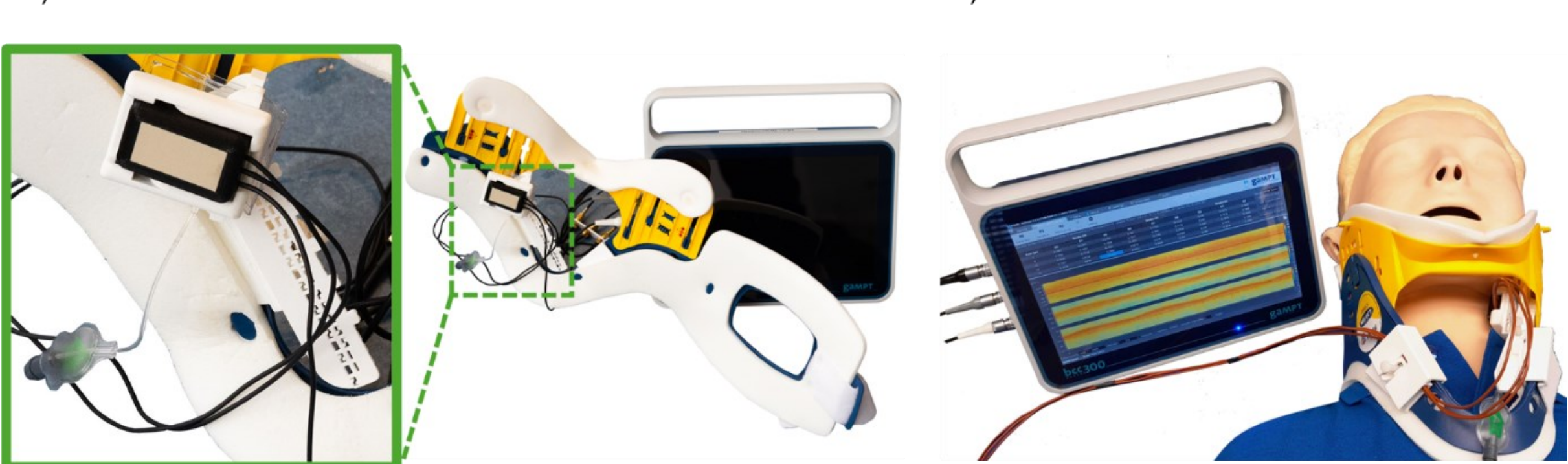

Fig. 1. Specially developed ultrasound system (GAMPT mbH) with Stifneck Select Collar™ (Laerdal) for sensor attachment; a) Left: US probe inside sensor mount on rail, fixed to the neck brace. Image is a cutout of the image area of the larger image to the right, cutout region marked with green lines. Right: System with open Stifneck Select Collar™ and inserted US-probe and the processing unit in the background; b) System applied to a dummy with simulated measurements.

In contemporary research and development, popular solutions for ultrasound probe fixation on the neck include adhesives and neckbands [8], [10], [11], [12], [13], [14], [15], [20], [21], [22]. Neckband constructions usually lack robust sensor skin contact, and the sensor possesses too many degrees of freedom during measurement. Sufficient contact adhesive solutions require firm adhesion to the skin. This increases the risk of skin trauma while placement corrections are not possible. Both solutions suffer from the problem that the head sensor relation is not fixed; movements will therefore cause increased signal noise and sensor vessel alignment loss.

To enable clinical and emergency personnel to perform hands-free blood flow measurements in the CCA during CPR as a feedback-system, we want to present a solution for a hands-free sensor probe attachment. The requirements originated in talks with medical experts and potential users of the emergency device and aim to provide a solution that features robust signal readings and high usability. In particular, the simplicity of the application should enforce resuscitation efforts in the setting of a cardiac arrest.

## II. Materials and Methods

### A. *Functional requirements*

In a collaborative process with medical experts of the Klinik und Poliklinik für Kardiologie, Universitätsklinikum Leipzig, physicists from the Gesellschaft für Angewandte Medizinische Physik und Technik (GAMPT mbH, Merseburg, Germany) and scientists from the Innovation Center Computer Assisted Surgery (ICCAS, Leipzig, Germany), a system for automated measurement of blood-flow in the ACC during resuscitation scenarios was proposed. The system should provide the following features:

- The device needs a complete hands-free approach for sensor attachment to the neck.
- After attaching the sensor, a fast detachment, repositioning and reattachment should be possible.
- Sensor positioning on the neck should be easily relatable to neck landmarks for documentation and comparable repositioning.
- Possible positions should cover most of the frontal neck area of one side (left or right).
- Attachment pressure should be applicable in an adjustable manner.
- No adhesive should be used to ensure sensor-skin-contact.

### B. *System setup*

Towards that purpose, we developed a new addition to an already existing neck brace system, the Laerdal Stifneck Select Collar™ (Laerdal Medical GmbH, Stavanger, Norway) (Fig. 1).

The final system for automatic acquisition of carotid blood-flow consists of a Stifneck Select Collar™, a custom 3D-printed sensor mount, a proprietary ultrasonic probe and a small computer with three ultrasonic control modules. The probe is fixed on a custom-made sensor mount, which in turn is fixed on the Stifneck Select Collar™, both shown in Fig. 1a. Fig. 1b shows the Stifneck Select Collar™ applied to a phantom for resuscitation exercises, the Resusci Anne® (Laerdal Medical GmbH, Puchheim, Germany). The sensor mount's components are the sensor holder, an inflatable pillow (TR-Band™, Terumo, Elkton, MD, USA), a rail and the rail holder. The sensor holder can fit a sensor with a base area of 25 mm x 41 mm. The sensor used in this setup has a height of 15 mm and two additional extensions at one end that fit under grooves in the holder and ensure the sensor does not fall out. The casing of the probe is made of disinfectable polymer with rounded corners and edges, to minimize risks of tissue damage.

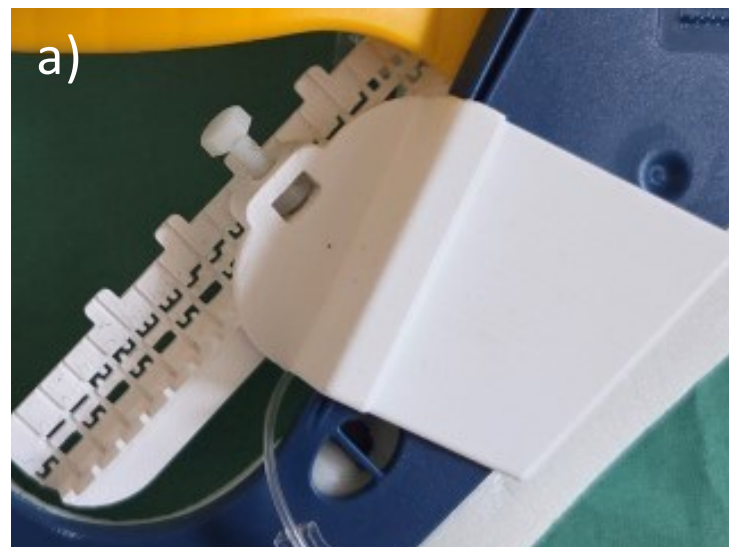

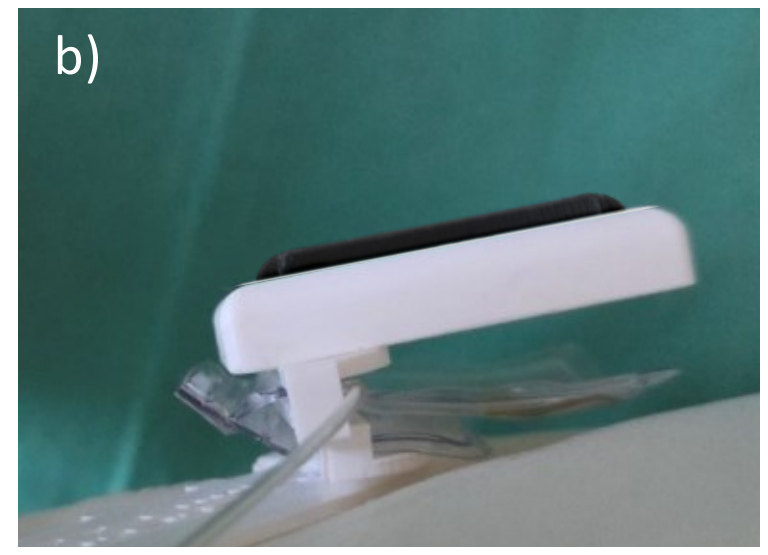

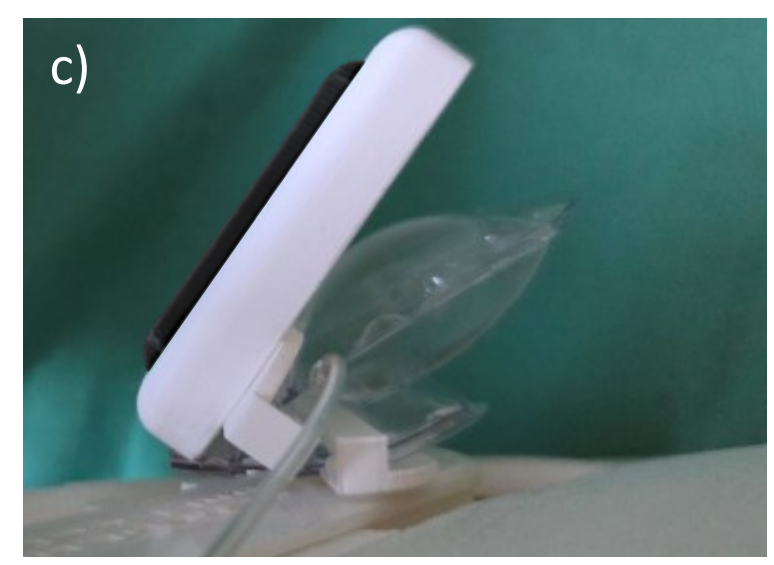

Fig. 2. Neck brace features for comparable repositioning and sensor pressure; a) sensor rail with notches for fixation on the sensor holder via extra bolt, 5 mm increments in notch difference allow for comparison of sensor position between patients and attachments; b) pillow (TR-Band™) for pressure build-up between sensor rail and sensor backside in deflated state; c) pillow (TR-Band™) in completely inflated state.

For security measures, the cables of the sensor probe are held by a cable relief opposite to the sensor mount, consisting of a strain relief bracket and the strain relief fixture. The sensor mount rail can be adjusted in its position, enabling the user to change the position of the sensor probe along the neck of the proband. To account for different-sized necks and adjust the necessary contact pressure, the pillow below the sensor holder can be inflated up to 18 ml, pressing the sensor probe towards the neck.

The ultrasonic probe used for measurements contains three different piezo-ceramics. Each ceramic has a different angle towards the probe surface, which ultimately connects to the proband's skin. The ultrasonic angles, results of the ceramic's orientation, toward the probe surfaces' normal are 25°, 0° and -12°. The probe is connected to the computer responsible for ultrasonic transmission and reflection measurement at each transducer ceramic. Besides the main computing unit, it features three ultrasound modules, one for each ceramic. The modules have synchronized communication, thus enabling the main unit to trigger an ultrasonic pulse via one ceramic by its respective ultrasound modules and measuring the response of targeted tissue via all ceramics/modules [23], [24].

The probe is secured with a specially designed holder printed from polylactic acid (PLA). The material is characterized by high stability and low water absorption and is resistant to most common disinfectants. For enabling repeatable measurements due to taking off the collar for medical interventions or changing the position of the sensor probe, the sensor can be adjusted along a rail (Fig. 2a). The rail features markings, enabling documentation of the sensor position in relation to the holding add-on, which is in a fixed place on the neck brace. To ensure sufficient contact between the skin and sensor surface, the sensor is pressed against the skin surface using a small, inflatable air cushion between the probe mount and sensor rail (Fig. 2b and Fig. 2c).

### *C. Study*

The study was conducted with healthy volunteers between 1st of August 2023 and 1st of March 2024. Study conductance was in accordance with the 1964 Declaration of Helsinki and received approval from the local ethics committee of the Leipzig University, Medical Faculty (reference number 183/23-ek). Volunteers provided written informed consent before participating in the study.

Hemodynamic stability and absence of cardiovascular symptoms were ensured at the time of measurement. Vital signs were recorded before the measurement. Exclusion from the study was limited mobility of the cervical spine, neck area injuries, or a known high-grade stenosis of the internal carotid artery.

During the study, the sensor contact surface was covered with ultrasound gel before each measurement. For measurements, the probands were in a supine position on an examination bed. After putting the Stifneck Select Collar™ around the proband's neck and securing the hold with a hook-and-loop fastener, the cushion was slowly inflated using a syringe, while a valve ensured that the air volume inside the cushion remained stable. Maximum volume was 18 ml, which limited the possibility of clamping off vessels or tissue damage due to excessive pressure. Afterwards, the position was documented, and the US device was used to check for audible and visual blood flow information using Doppler data. If verified, the position was used for an iterative measurement, the cushion was deflated, the rail bolt unsecured, and the rail moved to the next position. The bolt was fastened, and the pillow inflated again. Measurement and adjustment were repeated for as long as the sensor mount did not touch the lower jar.

### *D. Scaled assessments*

After execution of semi-automated measurement workflows, volunteers and study conductors (users) filled out questionnaires. Participants provided an estimation along a scale between 1 (lowest) and 10 (highest) to report how much pain (1 = none, 10 = intense pain) they felt due to the neck brace sensor mount combination and how comfortable (1 = very uncomfortable, 10 = very comfortable) the measurement was overall. The sensor attachment during measurements was assessed by the user and documented on a scale as well (1 = no stable connection, 10 = very robust connection).

### *E. Binary assessments*

During the measurement, the user evaluated whether a blood flow signal was audible and the flow curve on the real-time display of the device was usable. After measurement completion and removal of the neck brace, the user checked the measurement area for skin lesions as well. Signal audio, flow curve usability and lesion existence were documented with negative/positive checkboxes.

### *F. Statistical analysis*

Statistical analysis was done with MATLAB R2024a (The MathWorks, Inc., Natick, Massachusetts, United States). Descriptive analysis employed Matlab's functions for mean, standard deviation, median, minimum and maximum values. Tests for normal distributions were done with the Shapiro-Wilk test addon of Ahmed BenSaïda [25].

## III. Results

During the course of the study, 102 volunteers were recruited for the study's goal. Biological gender and relevant properties of study participants are listed in Table I.

TABLE I
OVERVIEW OF THE PARTICIPANTS' GENERAL AND NECK-RELATED CHARACTERISTICS

| Self-identified biological sex | diverse | female | male |
| --- | --- | --- | --- |
| Amount n [%] | 0 (0.0%) | 39 (38.2%) | 63 (61.8%) |

| Probands' characteristics | Mean | Std. | Median | Min / Max |
| --- | --- | --- | --- | --- |
| Age [years] | 37.7 | 16.8 | 33.0 | 19.0 / 83.0 |
| Height [cm] | 176.7 | 9.2 | 178.0 | 160.0 / 198.0 |
| Weight [kg] | 75.7 | 13.5 | 75.0 | 52.0 / 120.0 |
| Neck length [cm] | 18.6 | 1.9 | 18.0 | 16.0 / 23.0 |
| Neck circumference[cm] | 38.0 | 4.7 | 37.5 | 31.0 / 49.0 |
| BMI [kg/m²] | 24.2 | 3.6 | 23.53 | 17.2 / 34.4 |

Abbreviations: BMI: body mass index. Std.: standard deviation. Min.: minimum. Max.: maximum.

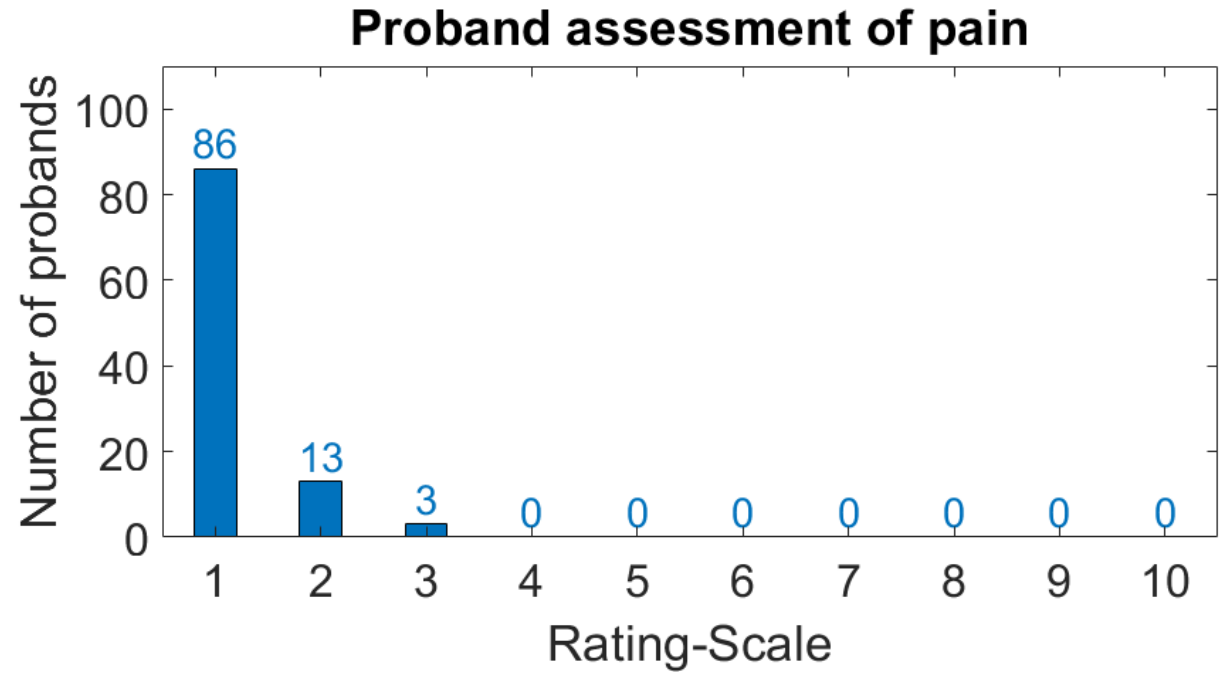

Fig. 3. Bar plot summarizing the results of the proband assessment regarding pain. The levels per metric range from 1 (minimal) to 10 (maximum). Pain (blue) is ranging from 1 to 3, with most people experience no pain or discomfort during the measurement.

Fig. 3 shows the bar plots containing the pain and comfort feelings that probands experienced, as well as the robustness of the evaluation of the sensor attachment, listed as support. More than 80% of the participants reported no pain during the measurement, selecting a score of 1 on the pain scale. Although volunteers were informed that they could stop the measurement and remove the neck brace - or have it taken off - at any time for any reason, none of them chose this option and completed the measurement.

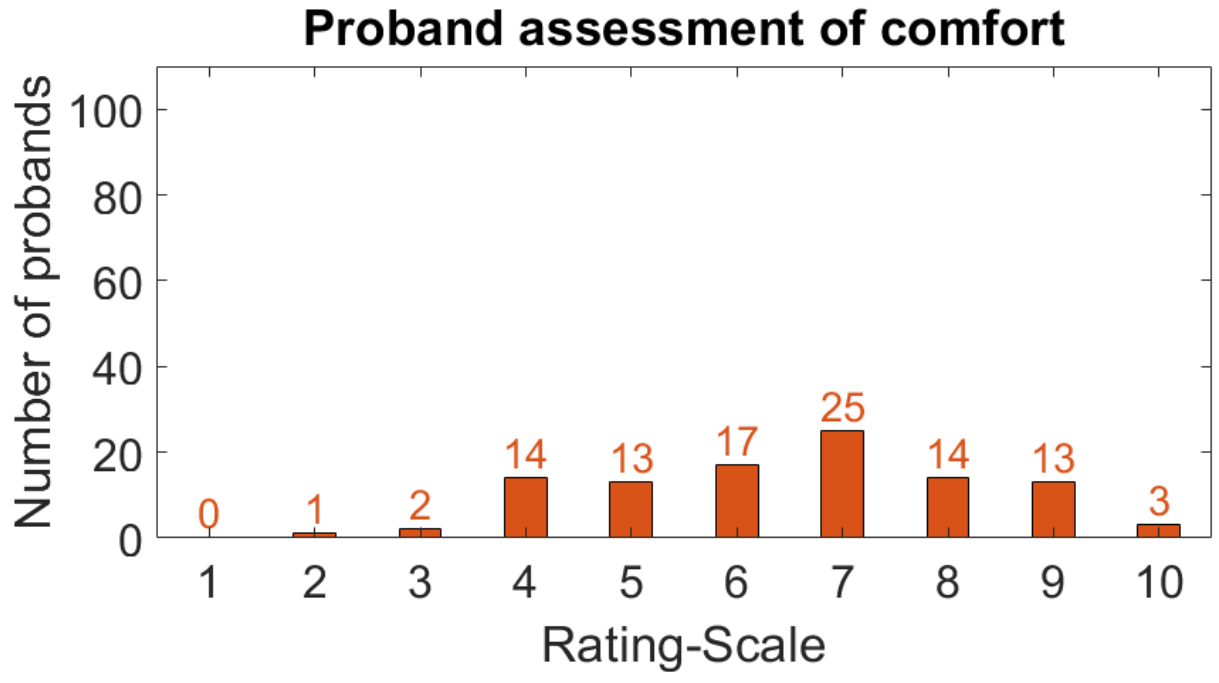

Fig. 4. Bar plot summarizing the results of the proband assessment regarding comfort. The levels per metric range from 1 (minimal) to 10 (maximum). Comfort (red) is ranging from 2 to 10 with most probands rating it at 7. On average the neck brace is therefore comfortable enough to allow for prolonged measurements without irritation.

Fig. 4 shows that neck brace comfort was more evenly spread around 7/10. Shapiro-Wilk test for normal distribution did not reject the null-hypothesis of the comfort results.

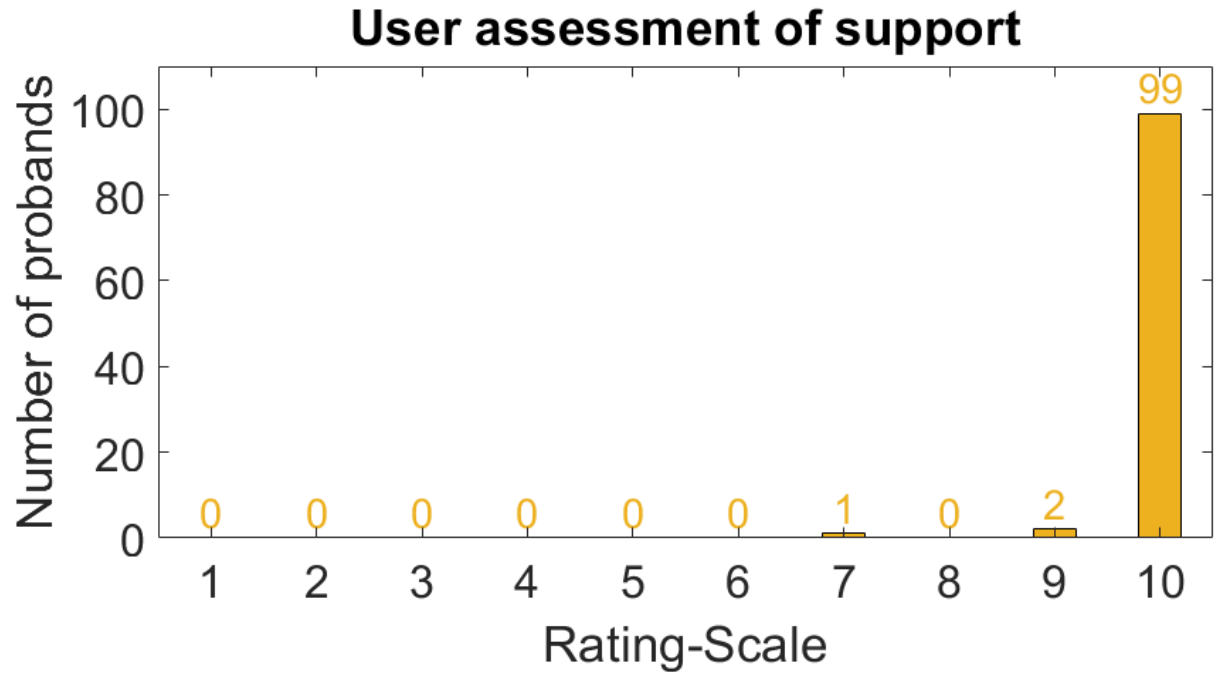

Fig. 5. Bar plot summarizing the results of the user assessment regarding support. The levels per metric range from 1 (minimal) to 10 (maximum). Support (yellow) is ranging from 7 to 10, with 99 of 102 cases showing best possible support, i.e., continous skin contact without displacement during measurement.

The user assessment of sensor attachment (support) is shown in Fig. 5 and was rated very robust, i.e., 10/10, by the user during 99 measurements. Twice the sensor attachment was designated less than very robust with 9/10, and only once the user selected 7/10, meaning attachment was still feasible, yet not achieved at the first try and subject to the volunteer's movements.

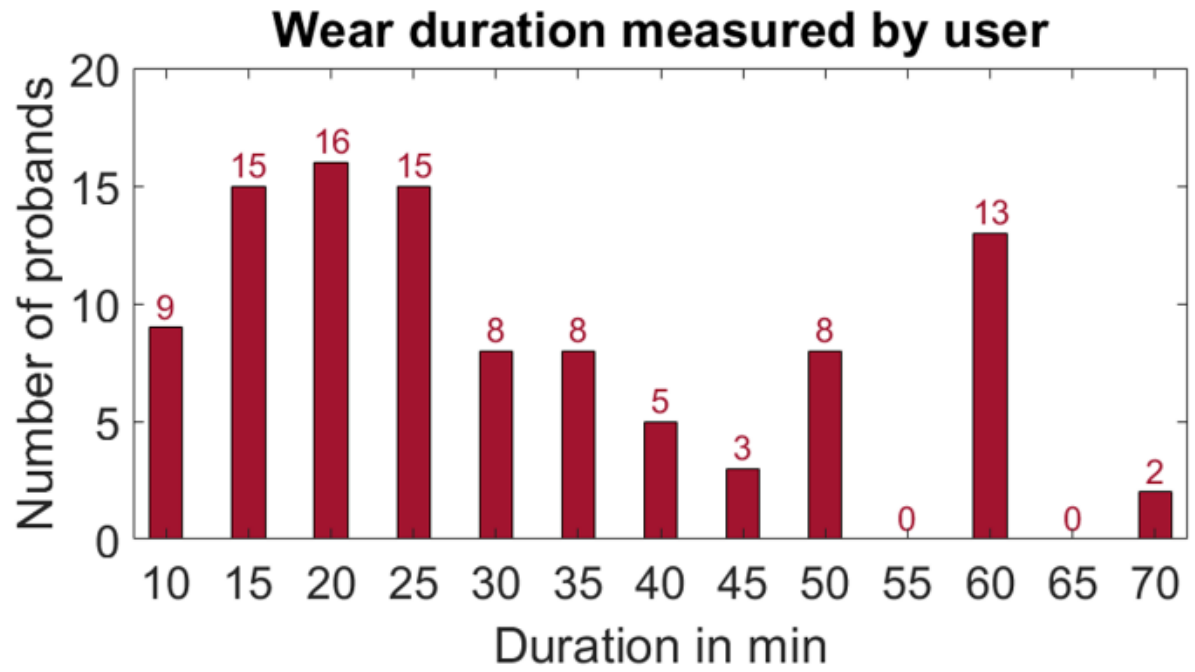

Fig. 6. Distribution of duration in minutes that the neck brace was worn; the duration in minutes is shown on the x-axis. It was rounded up to steps of five. The y-axis shows the number of occurrences with durations between 15 and 25 minutes being the three most occurring.

Fig. 6 shows the distribution of wearing duration across the measurements, ranging from 10 minutes of continuous neck brace application to a maximum of 70 minutes. The duration time was recorded and rounded up to the nearest 5-minute increment for statistical representation. Roughly 75% of the participants wore the neck brace for less than 50 minutes. A 60-minute wearing duration was achieved by 13 volunteers, and 2 volunteers experienced the maximum of a 70-minute duration. The 15 cases were reported to have experienced issues with the Windows operating system, requiring restarts and prolonging the measurement time. Table II provides a descriptive overview of the metrics pain, comfort, support and duration.

TABLE II
DESCRIPTIVE STATISTIC OVERVIEW OF ASSESSMENT METRICS AND NECK BRACE WEAR DURATION, ASSESSED BY PROBANDS (P) AND USERS (U)

| Metric | Pain (P) | Comfort (P) | Support (U) | Wear duration (U) |
|---|---|---|---|---|
| Average | 1.19 | 6.52 | 9.95 | 31.19 |
| Std. | 0.46 | 1.78 | 0.32 | 16.75 |
| Median | 1.00 | 7.00 | 10.00 | 25.00 |
| Minimum | 1.00 | 2.00 | 7.00 | 10.00 |
| Maximum | 3.00 | 10.00 | 10.00 | 70.00 |

Assessments with binary evaluation, provided by the expert users, are collected in Table III. Audible signals were observed during 94 of the 102 measurement sessions. The sensor probe attachment, therefore, satisfied the user(s) regarding signal stability and the continuous audio feedback caused by successful Doppler measurement. In 75/102 cases, the flow curve results provided by the custom US device allowed for qualitative flow analysis, according to the study conductors with medical backgrounds.

None of the participants showed signs of skin lesions after the measurement, and none contacted the study representative with any related issues since the measurement took place. Neither the pressure to maintain skin contact nor the prolonged wear of the neck brace caused any irritation or trauma that could be observed by the user(s) or felt by the participant.

TABLE III
DISTRIBUTION OF EVALUATION WITH BINARY CASES WHEN USING THE NECK BRACE, WHETHER AN AUDIBLE SIGNAL WAS ACHIEVED, AND WHETHER SKIN LESIONS WERE VISIBLE AFTER USE

| Metric | Audible Signal | Usable flow curve | Skin lesions |
|---|---|---|---|
| Positive \| Negative [n] | 94 \| 8 | 75 \| 27 | 0 \| 102 |
| Positive \| Negative [%] | 92.2 \| 7.8 | 73.5\|26.5 | 0 \| 100 |

### A. System safety

No reports indicated that the Stifneck Select Collar™ or probe attachment system negatively affected circulation and/or respiration.

## IV. DISCUSSION

The presented study examines the applicability of a hands-free, semi-automated ultrasound attachment device using a modified neck brace to enable safe and repeatable measurements of Doppler Flow velocity of the common carotid artery.

(1) A hands-free sensor attachment for a custom sensor ultrasound probe was constructed using the Stifneck Select Collar™.
(2) The device was easily applicable and was well tolerated without any safety concerns.
(3) Measurements of Doppler signals of the common carotid artery were feasible.

Hands-free measurements or wearable technology is already a popular topic among researchers, and a lot of valuable progress has been made in the past decades. Multiple research groups focus on hands-free solutions of ultrasound analysis, where sensor fixation is realized by adhesive materials [8], [10], [11], [13], [14], [15], [20], [21]. *Kenny et al.* presented their solution, FloPatch (FloSonics, Sudbury, Ontario, Canada), which is a continuous-wave (CW) Doppler device with two transducer arrays, capable of detecting any fluid movement below the surface the probe is attached to [10], [20]. The sensor can be applied to the skin via adhesive strips, enabling a hands-free measurement of carotid blood-flow once a proper position has been found. *Faldaas et al.* focused on a similar approach, with a device called *RescueDoppler*, which uses a two-transducer setup employing pulsed-wave (PW) Doppler and is applied with an adhesive patch as well, although it provides an additional guiding patch for placement support [11]. The PW Doppler enables the user to select a certain SV with available tissue depths of *RescueDoppler* reaching from 8 mm to 45 mm. Similarly, the *bioadhesive ultrasound* (BAUS) by *Wang et al.* features a small, rigid ultrasound probe, which provides ultrasonic measurements for up to 48 hours [21]. Alternatives to the patch-based solutions are wearables like braces, bands or necklaces, which have been proposed since 1999 by *Awad et al.,* among others [15], [22]. *Song et al.* developed a battery-powered neckband with US transducers and a Doppler module, capable of Bluetooth communication [12].

With adhesive solutions, if a sensor placement proved to be insufficient after attachment, a correction of the placement usually requires reattachment of the patches. In the CPR use case, this leads to lost resuscitation time, possible skin trauma and will affect the strength of the adhesive. Proper attachment and adequate pressure are therefore at risk, especially in wet conditions. Krüger et al. hinted at the shortcomings of adhesives in emergency scenarios, finding that correct placement could be difficult and skin adherence was poor in more than half of the patients [26]. They refer to studies involving biomimetic materials for better adhesion [27], [28], [29], yet admit these challenges appear unsolved in emergency settings [11], [26].

### A. The construction

The presented device addressed this issue with an approach that is not reliant on dry skin for good adhesion, allows for easy sensor removal and reattachment, and limits head movement. It has little injury risk since all rapid prototyping components feature smoothed edges and corners. Apart from the smoothed sensor probe, only the sensor case had skin contact if it was pushed until the lower jaw. Hook-and-loop fastener strips allowed a fast neck brace detachment in under two seconds, and the sensor rail and pressure pillow enable efficient sensor detachment, repositioning and reattachment, even with the neck brace on. The attachment pressure is manually adjustable and measurable via air volume or an additional barometer.

### B. The Application

Pain or any skin lesion, irritation, or trauma were never reported, which attributes to the safety and efficient combination of the sensor mount and the neck brace. Wear duration of up to 70 minutes provided no issues and didn't require any measurement termination, although a higher measurement duration originated in the device's operating system and has been addressed during the study. The combination of the Stifneck Select Collar™ with the sensor holder and the inflated cushion (TR-Band™, Terumo) for contact pressure did not cause any notable restriction in blood circulation or respiration either.

The system had extensive stability over time, which has been verified by medical users. Sensor fixation was resilient to moisture, e.g. sweat, and the relative alignment with the vessel had little risk of shifting, due to the fixed head position. Pressure adjustments or complete deflation are done fast and easily, to keep risks to a minimum. While neckbands are similar in their adjustability, they do not provide comparable stability and skin contact. Proper contact pressure of neckbands can cause breathing problems, occlude vessels and cause skin trauma, while straps and bands can interfere with invasive methods of medical personnel [12].

### C. The Measurement

Proband assessments of pain and comfort showed that the system was generally accepted and did not cause noticeable pain or intense discomfort. While some probands noted a slight discomfort associated with the pain grading 2 and 3, the majority felt nothing worth mentioning. The main concern is lower levels of comfort during the measurement, which were represented by the distribution of comfort ratings below level 7. The probands in question noticed the discomfort of the neck brace while resting their head on the bed. Considering the circumstances, the discomfort might be worth the advantages the system provides, since a) most patients of the OHCA use case will be unconscious, b) the risk of actual harm is low, and c) the advantages of CPR evaluation via a stabilized sensor attachment are far greater than potential discomfort after ROSC.

The system is less prone to placement errors due to the fixed mount on the Stifneck Select Collar™. The sensor rail allows only one degree of freedom, and preferred sensor positions can be configured quickly. Adhesive solutions require a complete certainty regarding the CCA position before sensor placement, an unlikely circumstance when facing a new patient. Neckbands are more easily corrected in their position, but the initial placement is just as complex. As long as no clear guidelines exist that define probable vessel depth and best sensor placement in relation to neck size, sensor position adjustments will most likely be necessary.

To our knowledge, no research group addressed the question of the optimal sensor placement depending on the patient anatomy conclusively. Finding the best spot is usually done by holding the patch and using US audio and visual feedback during blood flow times, requiring multiple people before adhesion [30], [31]. The RescueDoppler guiding patch shows the apparent optimal sensor placement in relation to the larynx, yet patient-specific anatomy could make this placement method invalid [26]. Formulating conclusive statements requires measurements with a sensor attachment method that provides a high degree of repeatability and reproducibility across different patients. Both solutions, adhesive and neckband, struggle with the realization of comparable results due to their respective disadvantages. Comparable sensor placements, quick probe adjustment, and stable measurement conditions are imperative and could be provided with the presented system.

### D. Limitations and future works

The proposed solution has been evaluated during a study for comparable US Doppler measurements with different neck positions on different probands. The largest limitation of the solution is the necessity to manually adjust the sensor probe position along the rail. This is due to the inconclusiveness of the current state-of-the-art regarding the best measurement position. The system, therefore, needs further research and development to draw connections between significant patients' characteristics and the neck position that is most likely to guarantee an alignment of US SV and the target vessel.

Forceful impacts during CPR applied in the medical use-case will add noise to the US measurement and increase the risk of SV misalignment again. The impact of those interferences on the measurement has to be researched more thoroughly. Displacement over time and hazard of sensor dislocation are minimized, but have yet to be evaluated under conditions comparable to the use case.

Additionally, under current circumstances, the pillow is held to the sensor rail at one point, creating a rotational axis during inflation, shown in Fig. 2. The resulting contact pressure between the sensor probe and skin might therefore be uneven, allowing gaps and corrupting US transmissions.

Finally, the used Stifneck Select Collar™ has come under critical review in recent years, since it partially covers neck areas that are important for intravenous injections. While the additional use of the Stifneck Select Collar™ as a platform for wearable sensor technology could increase appeal, the higher priority of easy access to jugular veins would require an alternate solution. An alteration to the current neck brace or a completely new solution would be required. The substitute would need to provide similar stability and usability, since most medical personnel are trained to use the Stifneck Select Collar™.

## V. CONCLUSION

In this work, a new sensor mount for a neck brace was presented and tested on healthy subjects. The sensor mount was designed for the use of ultrasonic measurements during out-of-hospital emergency scenarios and is used without adhesives. It enabled the user to quickly attach/detach a built-in US sensor to the neck for carotid blood-flow measurements, allowed for easy position adjustments, simple contact pressure regulation, and provided stable and repeatable sensor positioning. The system was generally accepted by the volunteers and caused no health damage. Further work includes the research of most likely sensor positions, testing alternative neck brace solutions for less obstruction of neck areas important for intravenous injections, and analyzing noise signals as well as usability under forceful impacts on phantoms.